\documentclass[fleqn,twoside]{article}
\usepackage{espcrc2}
\usepackage{graphicx}
\usepackage{bm}

\newcommand{\imag}{\mbox{i}\ }

\newcommand{\AmS}{{\protect\the\textfont2
  A\kern-.1667em\lower.5ex\hbox{M}\kern-.125emS}}

\title{Insight into nucleon structure from lattice calculations of moments of parton and generalized
  parton distributions}

\author{
J.W. Negele\address[MIT]{Center for Theoretical Physics,
        Massachusetts Institute of Technology,
 Cambridge, MA  02139, USA}{\thanks{speaker presenting this talk}},
R.C. Brower\address[UBOS]{Department of Physics,
Boston University,  Boston, MA  02215, USA}, P.
    Dreher\addressmark[MIT],
    R. Edwards\address[JLAB]{Thomas Jefferson National Accelerator Facility,      
    Newport News, VA  23606, USA}, 
    G. Fleming\addressmark[JLAB], Ph.
    H{\"a}gler\addressmark[MIT], 
            U.M. Heller\address[FLU]{American Physical Society, One Research Road, Ridge NY 11961-9000, USA}, \\
    Th. Lippert\address[UWUP]{Department of Physics, University of
    Wuppertal, D-42097 Wuppertal, Germany}, A.V.
    Pochinsky\addressmark[MIT], D.B.  Renner\addressmark[MIT], D.
    Richards\addressmark[JLAB], K.  Schilling\addressmark[UWUP], and W.
    Schroers\addressmark[MIT] }

\begin{document}

\begin{abstract}

This talk presents recent calculations in full QCD of the lowest three moments
 of generalized parton distributions and the insight they provide  into the behavior of nucleon electromagnetic form factors, the origin of the nucleon spin, and the transverse structure of the nucleon. In addition, new exploratory calculations in the chiral regime of full QCD are discussed.  \vspace{1pc}
\end{abstract}
\maketitle
%

%
%

\section{\label{sec:Introduction}INTRODUCTION}

Lattice field theory not only offers the prospect of precisely
calculating the experimentally observable properties of the nucleon
from first principles, but also offers the deeper opportunity of
obtaining insight into how QCD actually works in producing the rich
and complex structure of hadrons.  Beyond simply calculating numbers,
we would like to answer basic questions of hadron structure. For
example, how does the nucleon quark and gluon structure produce the
observed scaling behavior of form factors?  How does the
total spin of the nucleon arise from the spin and orbital angular
momentum of its quark and gluon constituents?  What is the transverse,
as well as longitudinal structure of the nucleon light-cone wave
function?  As the quark mass is continuously decreased from a world in
which the pion mass is 1 GeV to the physical world of light pions, how
does the physics of the quark model and adiabatic flux tube potentials
evolve into the physics of chiral symmetry breaking, where instantons,
quark zero modes, and the associated pion cloud play a dominant role?
As discussed below, contemporary lattice calculation are beginning to
provide insight into these and other fundamental questions in hadron
structure.

Because of asymptotic freedom, high energy lepton scattering provides precise measurements of matrix elements of the light-cone operator\\ [.3cm]
{\small
$
 {\cal O}(x) \!=\!\int \!\frac{d \lambda}{4 \pi} e^{i \lambda x} \bar
  \psi (\frac{-\lambda n}{2})\!\!
  \not n {\cal P} e\!^{-ig \int_{-\lambda / 2}^{\lambda / 2} d \alpha \, n
    \cdot A(\alpha n)}\!
  \psi(\frac{\lambda n}{2})
  $ } \\[.3cm]  
where $n$ is a unit vector along the light-cone.

The familiar quark distribution $q(x)$ specifying
the probability of finding a quark carrying a fraction $x$ of the
nucleon's momentum in the light cone frame is measured by the
diagonal nucleon matrix element, $ \langle P |{\cal O}(x) | P
\rangle = q(x) $.  Expanding  ${\cal O}(x) $ in local operators via the operator
product expansion generates the tower of twist-two
operators,
\begin{equation}
  \label{eq:gen-loc-curr}
  {\cal O}_q^{\lbrace\mu_1\mu_2\dots\mu_n\rbrace} = \overline{\psi}_q
   \gamma^{\lbrace\mu_1} \imag\stackrel{\leftrightarrow}{D}^{\mu_2} \dots
  \imag\stackrel{\leftrightarrow}{D}^{\mu_n\rbrace} \psi_q\,,
\end{equation}
and the diagonal matrix element $ \langle P | {\cal
  O}_q^{\lbrace\mu_1\mu_2\dots\mu_n\rbrace} | P \rangle$ specifies the
$(n-1)^{th}$ moment of the quark distribution $\int
dx\, x^{n-1} q(x) $.

The generalized parton distributions  $ H(x, \xi, t)$ and  $ E(x, \xi, t)$  \cite{Muller:1994fv,Ji:1997ek,Radyushkin:1997ki,Diehl:2003ny} are
measured by off-diagonal matrix elements of the light-cone operator
\begin{eqnarray}
\langle P' |{\cal O}(x) | P \rangle\! &=&  \!\langle\!\langle \not n \rangle\!\rangle
H(x, \xi, t) \nonumber  \\ &+&  \frac{i \Delta_\nu} {2 m}  \langle\!\langle \sigma^{\alpha \nu} n_{\alpha} \rangle\!\rangle
E(x, \xi, t),
\end{eqnarray} 
where
$\Delta^\mu = P'^\mu - P^\mu$, $ t  = \Delta^2$, $\xi = -n \cdot \Delta /2$, and
$\langle \!
\langle \Gamma \rangle \! \rangle = \bar U(P') \Gamma U(P)$ for Dirac spinor $U$.
Off-diagonal matrix
elements of the tower of twist-two operators
$ \langle P' | {\cal
  O}_q^{\lbrace\mu_1\mu_2\dots\mu_n\rbrace} | P \rangle$ yield moments of the
generalized parton distributions, which in the special case of $\xi$ =
0, are
\begin{eqnarray}
 \int dx\, x^{n-1} H(x, 0, t) & = &   A_{n, 0}(t) \nonumber \\
 \int dx\, x^{n-1} E(x, 0, t) &=&  B_{n, 0}(t),
 \label{gff1}
\end{eqnarray}
where  $ A_{n, i}(t)$ and $B_{n, i}(t)$ are referred to
as generalized form factors (GFF's).
Analogous expressions in which the light-cone operator $\mathcal{O}(x)$ and
twist-two operators contain an additional $\gamma_5$ measure the
longitudinal spin density, $\Delta q(x)$ and spin-dependent generalized
parton distributions $\tilde H(x, \xi, t) $ and $\tilde E(x, \xi, t) $ with moments $\tilde A_{n, i}(t)
$ and $\tilde B_{n, i}(t)$. 

In this talk, I will discuss recent calculations \cite{Hagler:2003jd,Negele:2003ma,Schroers:2003mf,Haegler:2003pr} of the generalized
form factors $A_{(n=1,2,3), 0}(t)$ and $\tilde A_{(n=1,2,3), 0}(t)$ in full,
unquenched QCD in the currently computationally accessible domain that I  refer to as the ``heavy pion world'' and discuss their physical significance. In addition, I will discuss initial efforts to explore the chiral regime in which the pion mass is sufficiently light that one can use chiral perturbation theory to extrapolate to the physical pion mass. Although results in the heavy pion world cannot be directly compared with experiment, they nevertheless provide important insight into how QCD works, and provide the first step in the ultimate program of studying how hadronic physics evolves from the heavy pion world to our physical world.



\pagebreak

\section{\label{sec:lattice-calculation}LATTICE CALCULATION}

The lowest three moments of spin-independent GPD's considered in this talk are
{
\begin{eqnarray}
  \label{eq:explicit}
 \lefteqn{ \langle P' | {\cal O}^{\mu_1} | P \rangle  \!=\!
  \langle \! \langle \gamma^{\mu_1 }\rangle \! \rangle A_{10}(t)
  \nonumber} \\
  &+&\! \frac{\imag}{2 m} \langle \! \langle \sigma^{\mu_1
    \alpha} \rangle \! \rangle
  \Delta_{\alpha} B_{1
    0}(t)\,, \nonumber \\ [.5cm]
 \lefteqn{ \langle P' | {\cal O}^{\lbrace \mu_1 \mu_2\rbrace} | P \rangle  \!=\!
  \bar P^{\lbrace\mu_1}\langle \! \langle
  \gamma^{\mu_2\rbrace}\rangle  \! \rangle
  A_{20}(t) \nonumber }\\
  &+&\! \frac{\imag}{2 m} \bar P^{\lbrace\mu_1} \langle \! \langle
  \sigma^{\mu_2\rbrace\alpha}\rangle \! \rangle \Delta_{\alpha} B_{2 0}(t)
  \nonumber \\
  &+&\!\frac{1}{m}\Delta^{\{ \mu_1}   \Delta^{ \mu_2 \} }
  C_{20}(t)\,, \nonumber \\[.5cm]
 \lefteqn{ \langle P' | {\cal O}^{\lbrace\mu_1 \mu_2 \mu_3\rbrace} | P \rangle
  \!=\! \bar P^{\lbrace\mu_1}\bar P^{\mu_2} \langle \! \langle
  \gamma^{\mu_3\rbrace}
  \rangle \! \rangle A_{30}(t) \nonumber }\\
  &+&\! \frac{\imag}{2 m} \bar P^{\lbrace \mu_1}\bar P^{\mu_2}
  \langle \! \langle \sigma^{\mu_3\rbrace\alpha} \rangle \! \rangle
  \Delta_{\alpha} B_{3 0}(t) \nonumber
  \\
  &+&\! \Delta^{\lbrace \mu_1}\Delta^{\mu_2} \langle \! \langle
  \gamma^{\mu_3\rbrace}\rangle \! \rangle A_{32}(t) \nonumber \\
  &+&\! \frac{\imag}{2 m} \Delta^{\lbrace\mu_1}\!\Delta^{\mu_2}
  \langle \! \langle \sigma^{\mu_3\rbrace\alpha}\rangle \! \rangle
  \Delta_{\alpha} B_{3 2}(t), 
\end{eqnarray}
}

Generalized form factors $A_{(n=1,2,3),0}(t)$ and $\tilde A_{(n=1,2,3),0}(t)$ were calculated using the new method introduced in Ref. \cite{Hagler:2003jd}. 
We considered all the combinations of $\vec P$ and $\vec P'$ that could produce the same four-momentum transfer $t=(P'-P)^2 $, subject to the conditions that 
$\vec P = \frac{2\pi}{a N_s}(n_x,n_y,n_z)$ and
$\vec P'=(0,0,0)$ or $\frac{2\pi}{a N_s}(-1,0,0)$. Using all these momentum combinations  for a given $t$ below 3.5 GeV, we calculated all the $H(4)$ cubic group lattice operators and index combinations producing the same continuum GFF's, obtaining an overdetermined set of equations from which we extracted the most statistically accurate measurement of the desired GFF's the available lattice data can provide. As discussed in connection with Fig.~\ref{fig:method}, the errors are substantially smaller than obtained by the common practice of measuring a single operator with a single momentum combination.


 
We calculated connected diagram contributions using  approximately 200 
SESAM \cite{Eicker:1998sy} full QCD configurations with Wilson fermions at $\beta = 5.6$ on $16^3 \times 32$ lattices. These calculations in the heavy pion world were performed at each of three quark masses, $\kappa$= 0.1570, 0.1565, and 0.1560, corresponding to pion masses defined by $r_0$ of 744, 831, and 897 MeV respectively.

 
 \section{\label{sec:electromagnetic-form-factors}ELECTROMAGNETIC FORM \\ FACTORS}

                \begin{figure}[t]
                 \vspace*{-0.8cm}
                  \includegraphics[clip=false,scale=0.29]{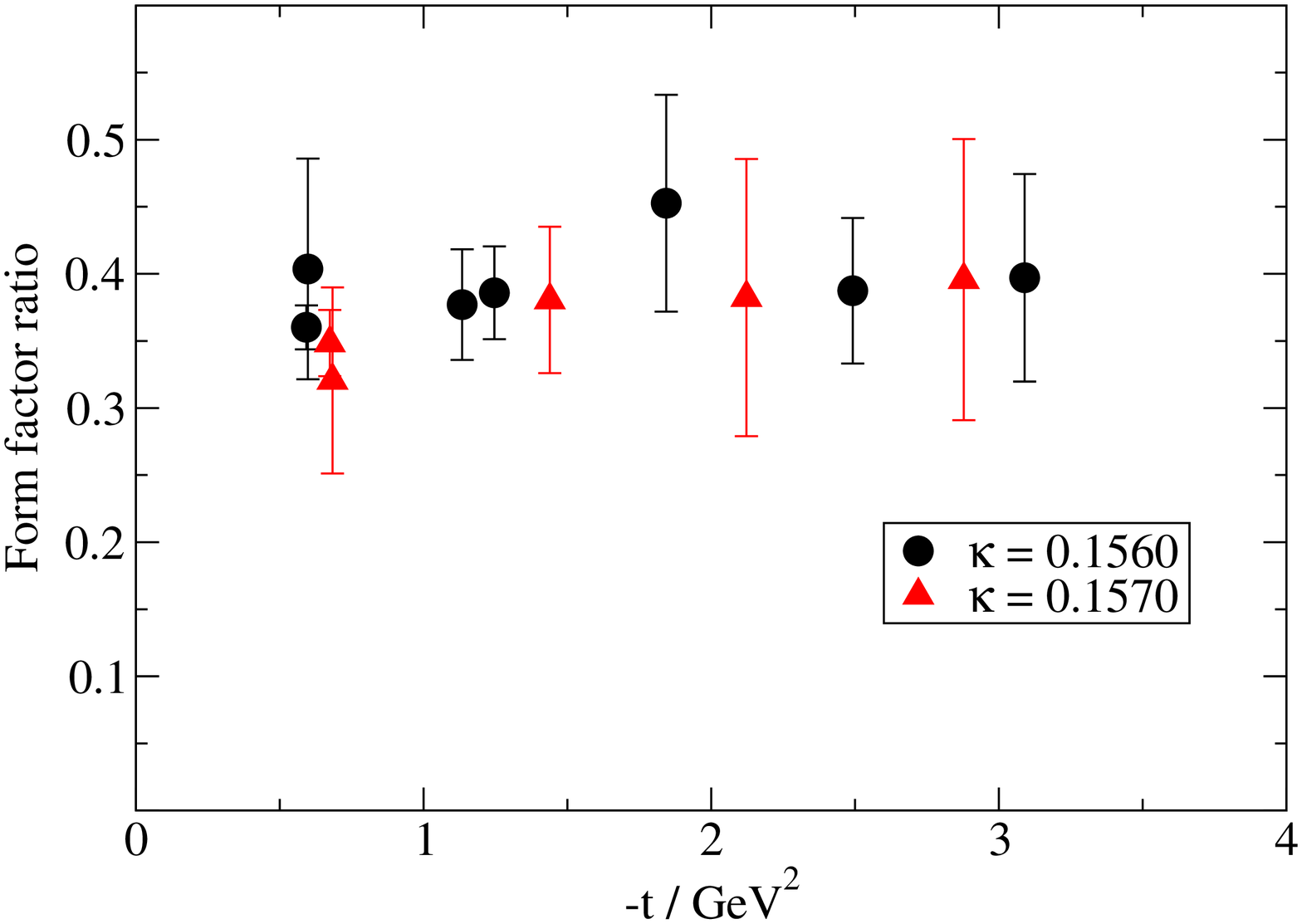}
                    \includegraphics[clip=false,scale=0.60]{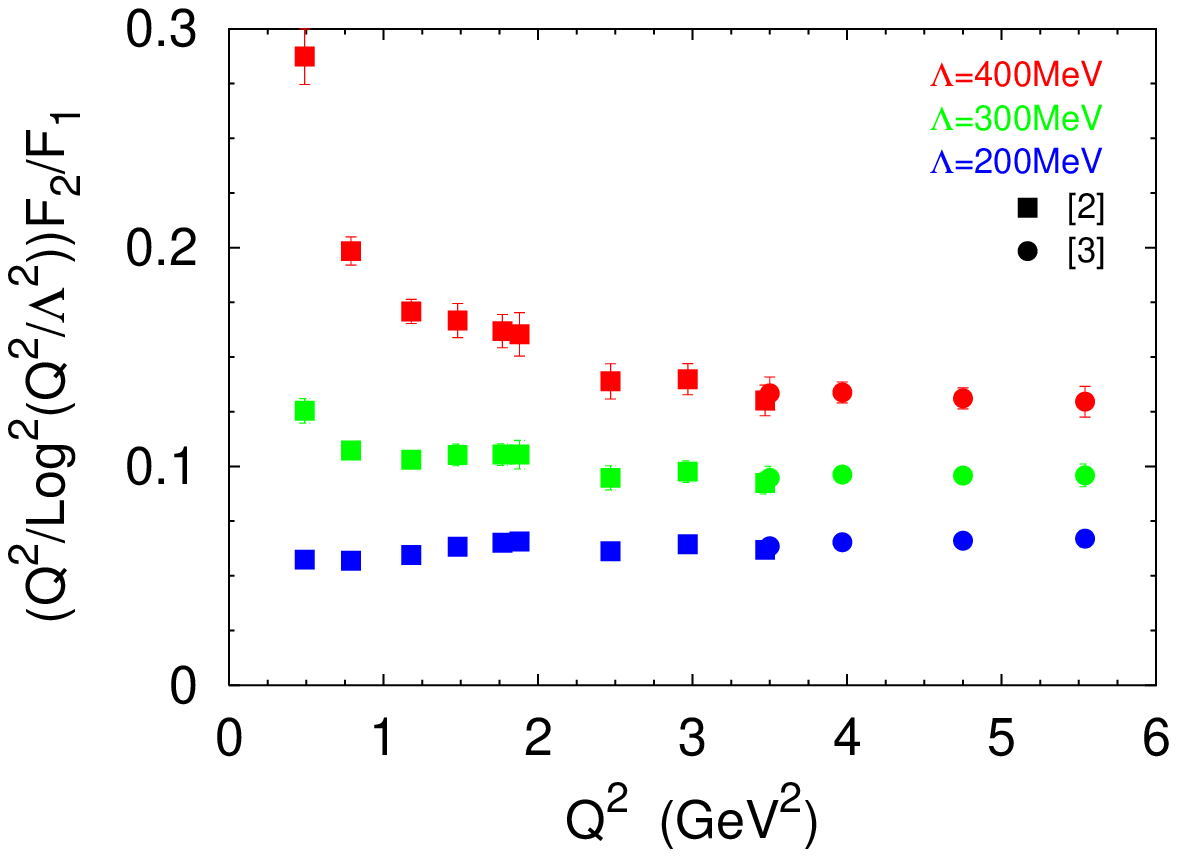}
          \vspace*{-1.0cm}
                  \caption{Electromagnetic form factor ratio 
                    $\frac{Q^2 F_2(Q^2)}{\log^2 (Q^2 / \Lambda^2)
                      F_1(Q^2)}$. The upper plot shows lattice results
                    plotted with $\Lambda$=300 MeV. Experimental data
                    taken from Ref.~\cite{Belitsky:2002kj} are shown
                    below, and should be multiplied by the anomalous
                    magnetic moment, $\kappa$ = 1.79, for comparison
                    with the lattice results.}
                  \label{fig:ji}
                  \vspace*{-.5cm}
                \end{figure}

One of the early successes of perturbative QCD was the understanding
of how the short range quark structure of a hadron governs the
behavior of exclusive processes at large momentum transfer. However,
whereas simple counting rules suggested that $F_2 \sim F_1 / Q^2$,
experimental data from JLab \cite{Gayou:2001qd} shows that $F_2$ falls off much
more slowly. Theoretically, it has recently been shown
\cite{Belitsky:2002kj} that the next to leading order light cone wave
function yields $F_2 \sim F_1 {\log^2 (Q^2 / \Lambda^2)} / {Q^2}$,
and the agreement between this prediction and the JLab data is shown
in the lower portion of Fig.~\ref{fig:ji}.
 
 Since the short range quark structure dominates this physics, it is reasonable to expect that omission of the pion cloud in the heavy pion world should not destroy the qualitative behavior.  Indeed, our lattice results plotted  in the top portion of Fig.~\ref{fig:ji} for the value $\Lambda$ = 0.3 GeV yields excellent agreement with the $Q^2$ behavior of the experimental data.

 \section{\label{sec:origin-of-the-nucleon's-spin}ORIGIN OF THE NUCLEON'S SPIN}

\begin{figure}[t]
 \vspace*{-0.75cm}
     \includegraphics[scale=0.30,clip=true,angle=270]{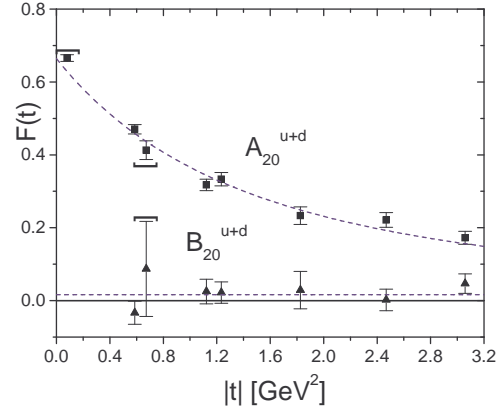}
      \vspace*{-0.8cm}
  \caption{Generalized form factors $A^{\mbox{\tiny
    u+d}}_{20}(t)$ and $B^{\mbox{\tiny
    u+d}}_{20}(t)$, with dipole fits denoted by dashed curves. }
  \label{fig:A2B2}
   \vspace*{-.5cm}
\end{figure}

\begin{figure}[t]
 \vspace*{-0.75cm}
   \includegraphics[scale=0.30,clip=true,angle=270]{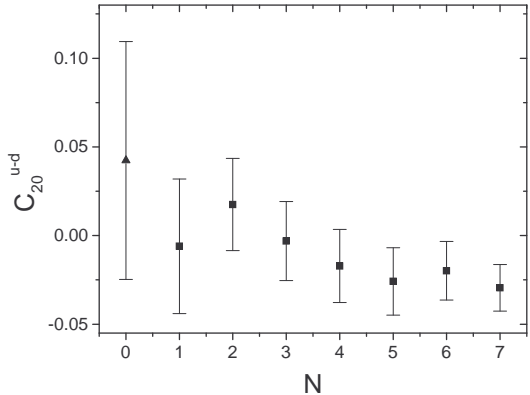}
      \vspace*{-0.4cm}
  \caption{Example of the effectiveness of the overdetermined fit for $C^{\mbox{\tiny
    u-d}}_{20}(t)$. The N=0 point uses three operators at a single external momentum combination to determine three form factors. The remaining points use six operators and N external momentum combinations. }
  \label{fig:method}
   \vspace*{-.5cm}
\end{figure}

\begin{figure}[t]
 \vspace*{-1.7cm}
  \hspace*{-1.3cm}     \includegraphics[scale=0.41,clip=true,angle=0]{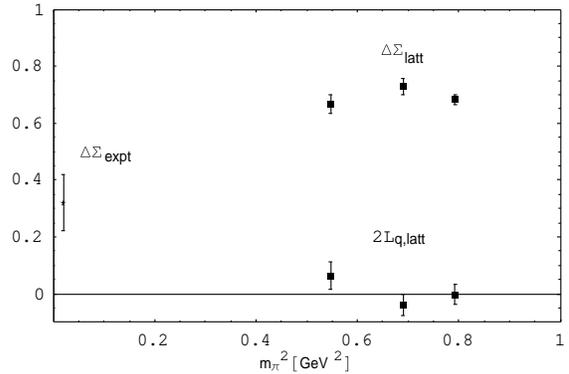}
      \vspace*{-2.0cm}
  \caption{Fraction of the proton spin arising from the quark spin, $\Delta \Sigma$, and from the quark orbital angular momentum, $L_q$, as described in the text. }
  \label{fig:quark_ang_mom}
   \vspace*{-.5cm}
\end{figure}

In the nonrelativistic quark  model with three quarks in angular momentum zero single particle states, the total proton spin of $1/2$ arises trivially from adding the spins of the three quarks. The so-called spin crises arose when deep inelastic scattering measurements of the lowest moment of the spin-dependent structure function, $\Delta \Sigma = \langle 1 \rangle_{\Delta u} + \langle 1 \rangle_{\Delta d}$, indicated that only of the order of 30\% of the nucleon spin arises from quark spins. Physically,  in the heavy pion world where the quarks become less relativistic and are described by the quark model, it is reasonable to expect most of the nucleon spin to arise from the valence quark spin. As the quarks become lighter, one expects this fraction to decrease as the relativistic quarks acquire more angular momentum and instanton effects, for example, remove helicity from the valence quarks and transfer it to the gluons and to quark-antiquark pairs. We would hope to observe this behavior clearly on the lattice. 
 
The total quark contribution to the nucleon spin is given by the extrapolation to $t=0$ of  $A^{\mbox{\tiny u+d}}_{20}(t)$ and $B^{\mbox{\tiny  u+d}}_{20}(t)$ shown in Figure 2.  Since  $A^{\mbox{\tiny u+d}}_{20}(t)$ is calculated directly at   $t=0$  and  $B^{\mbox{\tiny  u+d}}_{20}(t)$ is well fit by a constant that is measured to be nearly zero with small errors, the connected contribution to the angular momentum is measured to within a few percent. 

The calculation of the generalized form factors $A_{20}(t)$, $B_{20}(t)$, and $C_{20}(t)$
provides an excellent example of the power of the new  method introduced in Ref.~\cite{Hagler:2003jd}. Figure~\ref{fig:method} shows how adding additional operators and external momentum combinations reduces the error in $C_{20}(t)$ by a factor of 5 from the standard minimally determined case to our highly overdetermined case.

Combined with the results of $\Sigma$ from reference \cite{Dolgov:2002zm}, we obtain the connected diagram contributions to the decomposition of nucleon spin shown in Tab.~\ref{tab:spin} and plotted in Fig.~\ref{fig:quark_ang_mom}. Similar results have been obtained in Refs.~\cite{Gockeler:2003jf,Mathur:1999uf}.      To the extent that the disconnected diagrams do not change the qualitative behavior, we conclude that of the order of 70\% of the spin of the nucleon arises from the quark spin and a negligible fraction arises from the quark orbital angular momentum in a heavy pion world where $m_\pi \sim$ 700 - 900 MeV. This behavior is just as expected from the arguments above.
As lattice calculations approach the chiral limit, it will be interesting to fill in this graph and observe the quark spin contribution decrease to  $\sim$ 30\% to agree with experiment.

\begin{table}[t]
  \begin{tabular}[b]{*{3}{c|}c}
    \hline
         $\kappa $& 0.1570 & 0.1565 &  0.1560 \\ \hline
          $\Delta \Sigma$ &0.67$\pm$ .04 & 0.73 $\pm$ .03 & 0.68 $\pm$ .02\\ \hline   
          $2 L_q$ &0.06$\pm$ .05 & -0.04 $\pm$ .04 & 0.00 $\pm$ .03\\ \hline
          $2 J_q$ &0.73$\pm$ .04 & 0.69$\pm$ .02 & 0.68 $\pm$ .03\\ \hline
    \end{tabular}
  \caption{Fraction of nucleon spin arising from quark spin, $\Delta \Sigma$, quark orbital angular momentum,  $2 L_q$, and quark total angular momentum,  $2 J_q.$ }
  \label{tab:spin}
\end{table}

 \section{\label{sec:transverse-structure-of-the-nucleon}TRANSVERSE STRUCTURE OF THE NUCLEON}
 
\begin{figure}[t]
  \begin{tabular}{*{3}{c}}
  \hspace{-.3cm}
    \includegraphics[width=0.46\textwidth,clip=true,angle=0]{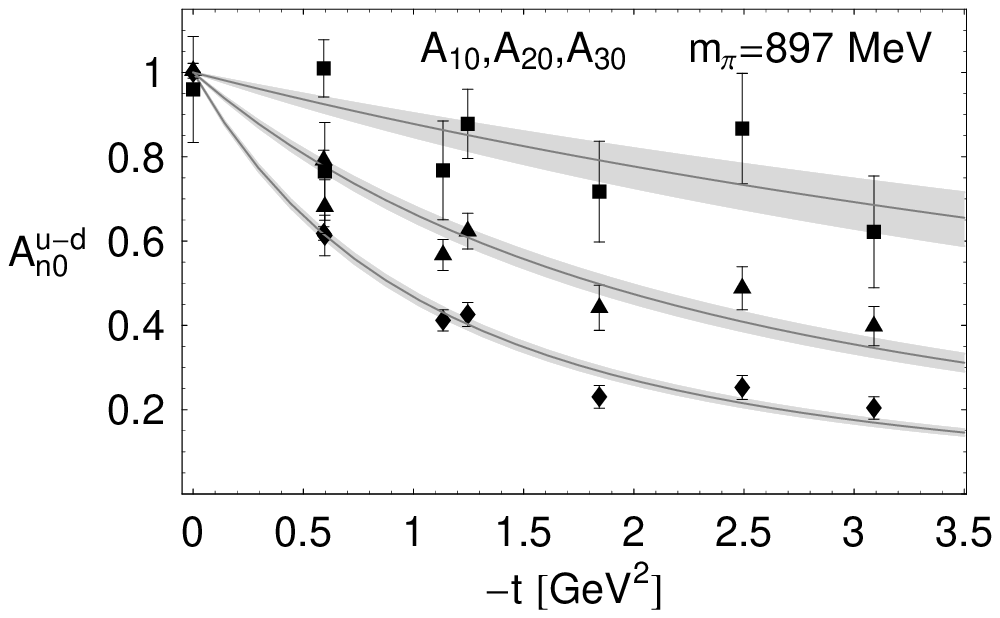}\\
\hspace{-.3cm}
    \includegraphics[width=0.46\textwidth,clip=true,angle=0]{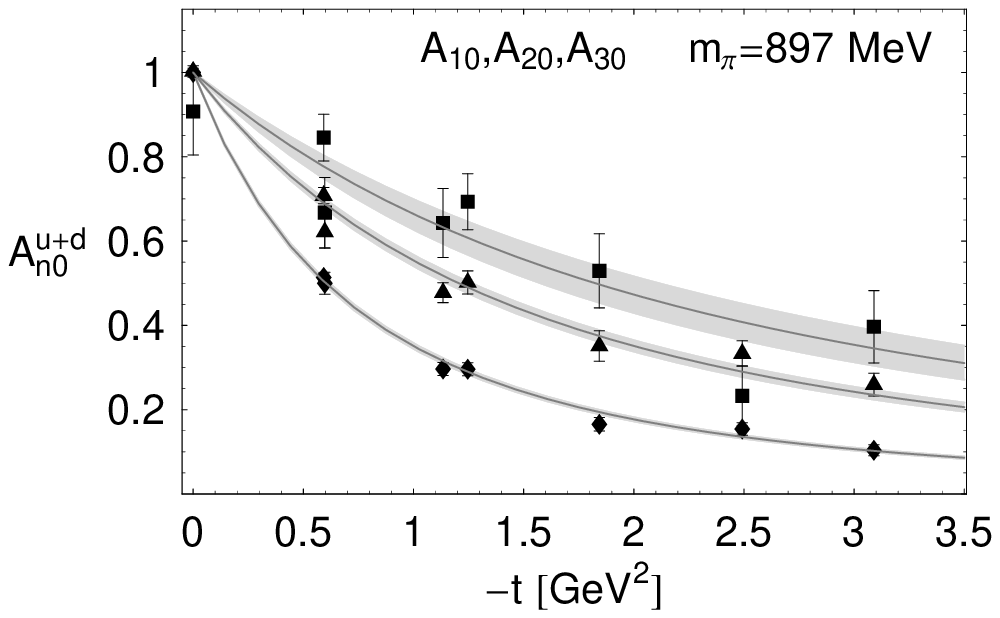}\\
  \end{tabular}
\caption{ Normalized generalized 
  form factors $A^{\mbox{\tiny u-d}}_{n,0}(t)$ 
  and $A^{\mbox{\tiny u+d}}_{n,0}(t)$    for n=1 (diamonds), n=2 (triangles) and n=3 (squares).  }

 \label{fig:slopes1}
\end{figure}

\begin{figure}[t]
  \begin{tabular}{*{3}{c}}
  \hspace{-.3cm}
   \includegraphics[width=0.46\textwidth,clip=true,angle=0]{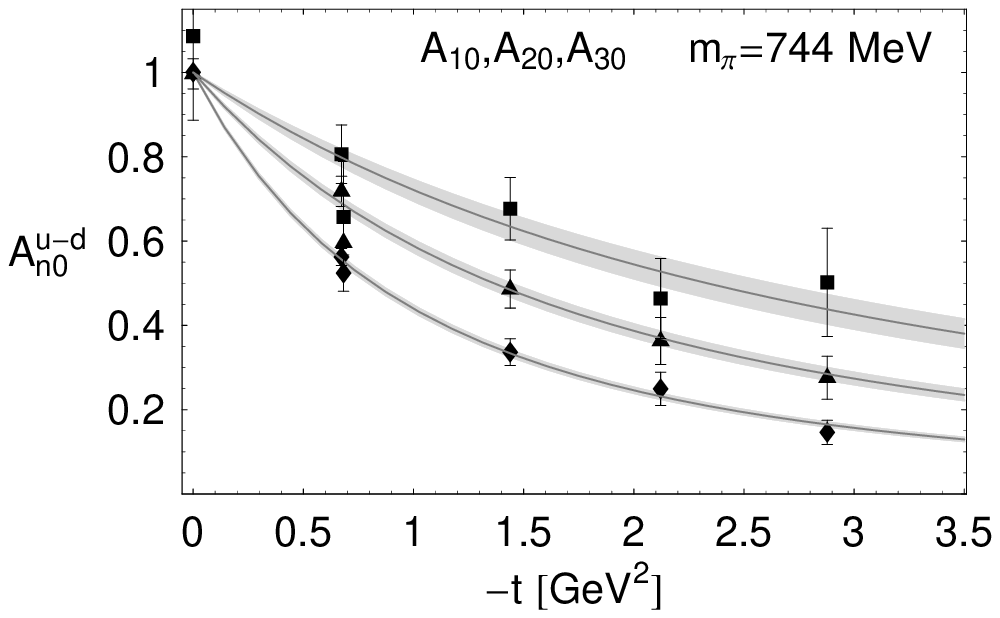}\\
\hspace{-.3cm}
  \includegraphics[width=0.46\textwidth,clip=true,angle=0]{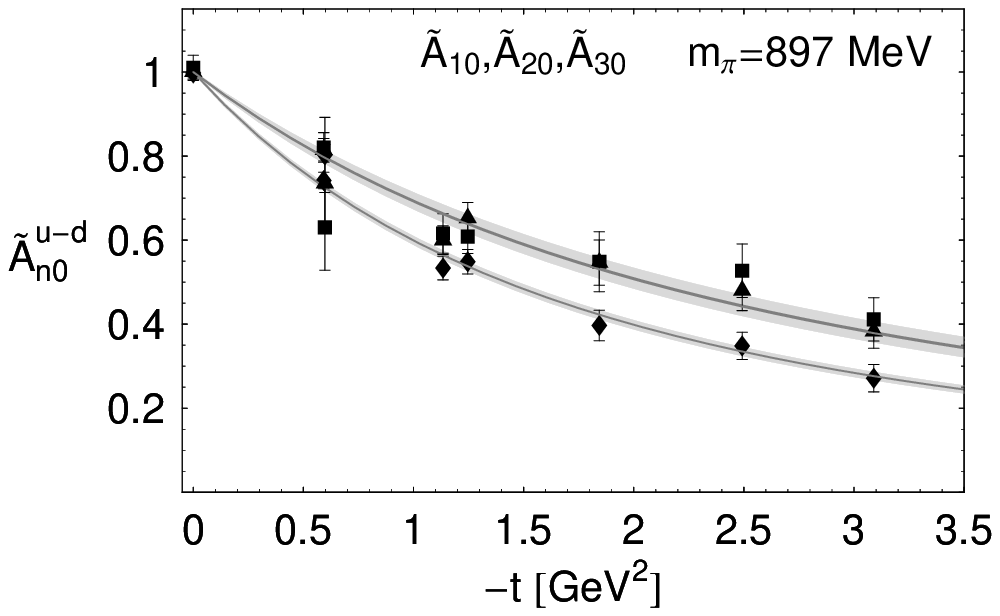}  
  \end{tabular}
\caption{ Normalized generalized 
  form factors $A^{\mbox{\tiny u-d}}_{n,0}(t)$ 
  and $\tilde A^{\mbox{\tiny u-d}}_{n,0}(t)$    for n=1 (diamonds), n=2 (triangles) and n=3 (squares).  }
   \label{fig:slopes2}
   \end{figure}

\begin{figure}[t]
 \vspace*{0.1cm}
    \includegraphics[scale=0.3,clip=true,angle=0]{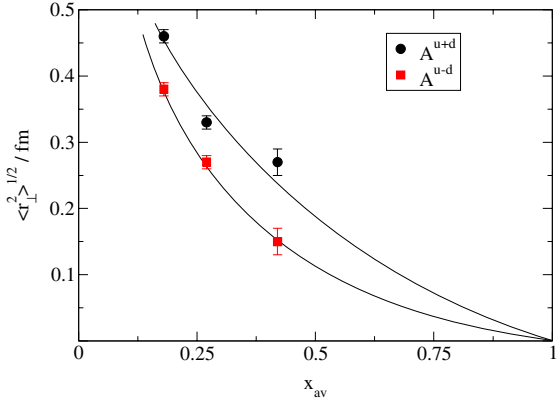}
       \vspace*{-0.5cm}
  \caption{Transverse rms radius of the proton light cone wave
       function as a function of the average quark momentum fraction,
       $x_{\mbox{\tiny av}}$, for each measured moment. }
  \label{fig:rms}
   \vspace*{-.5cm}
\end{figure}

In general, $ H(x, \xi, t)$ is complicated to interpret physically because it
combines features  of both parton distributions and form factors, and depends on
three kinematical variables: the momentum fraction $x$, the longitudinal
component of the momentum transfer $\xi$, and the total momentum transfer
squared, $t$.  In the particular case in which $\xi = 0$, however, Burkardt
\cite{Burkardt:2000za} has shown that  $ H(x, 0, t)$, as well as its spin-dependent
counterpart  $ \tilde H(x, 0, t)$, has a simple and revealing physical
interpretation.

It is useful to consider a mixed representation in which transverse coordinates
are specified in coordinate space, the longitudinal coordinate is specified in
momentum space and one uses light-cone coordinates for the longitudinal and time
directions: $x^{\pm} = (x^0 \pm x^3)/ \sqrt{2}$, $p^{\pm} = (p^0 \pm p^3)/
\sqrt{2}$.
Using these variables, letting $x$ denote the momentum fraction and 
$b_\perp$ denote the transverse displacement (or impact parameter) of the light
cone operator relative to the proton state,  one may define an impact parameter
dependent parton distribution
\begin{eqnarray*}
q(x,b_{\perp})\! \equiv \! \langle P^+\!, R_{\perp}\!\!=0, \lambda | \hat {\cal O}
(x,b_\perp) |P^+\!, R_{\perp}\!\!=0, \lambda \rangle\,, 
\end{eqnarray*}
where
\[
{\cal  O}_q(x,b_\perp)\! = \!\!\int \! \frac{d x^-}{4 \pi} e^{i x p^+ x^-} \!\! \!\bar
  q (\!-\frac{x^-}{2}, b_\perp)
 \gamma^+\!
  q(\frac{x^-}{2}, b_\perp).
\]

Burkardt shows that the generalized parton
distribution $H(x, 0, t)$ is the Fourier transform of the impact
parameter dependent parton distribution, so that
        \begin{eqnarray}
H(x, 0, -\Delta_\perp^2) \!&=&\! \int d^2 b_\perp q(x, b_{\perp})\,e^{
        i \vec b_\perp \cdot  \vec \Delta_\perp} \nonumber \\
        A_{n,0}( -\Delta_\perp^2)  
        \!&=&\! \int d^2 b_\perp \int dx\, x^{n-1} \nonumber \\
        && \qquad\times q(x, b_{\perp})\,e^{
        i \vec b_\perp \cdot  \vec \Delta_\perp}\!,
        \end{eqnarray}
where the second form follows from Eq.~(\ref{gff1}).
Although one normally only expects a form factor to reduce to a Fourier
transform of a density in the non-relativistic limit, Ref.~\cite{Burkardt:2000za} shows
that special features of the light cone frame also produce this simple result in
relativistic field theory.
Thus, if one could  measure $H(x, 0, t)$ as a
function of $ t$, one could determine how the transverse distribution of
partons varies with momentum fraction $x$.


Physically, we expect the transverse size of the nucleon to depend
significantly on the momentum fraction $x$. Averaging $q(x,b_\perp)$ over all $x$,
which produces $A_{1,0}(t)$ and thus corresponds to calculating the form factor, the size is characterized by the transverse rms radius $  \langle r_\perp^2 \rangle^{\frac{1}{2}}  =  \langle x_1^2 + x_2^2
\rangle^{\frac{1}{2}} = \sqrt{{\frac{2}{3}}} \langle r^2 \rangle^{\frac{1}{2}}$.
From the experimental electromagnetic form factor, the transverse rms charge radius of the  proton is 0.72 fm.  As $x \to 1$, the active parton carries
all the momentum and the spectator partons give a negligible contribution.
In this case the active parton represents the (transverse) center of momentum, and the
distribution in impact parameter reduces to a 
delta function $\delta(b_{\perp}) $ with zero spatial extent.
Indeed, explicit light cone wave functions \cite{Brodsky:2000xy,Diehl:2000xz} bear out
this expectation, with the result \cite{Diehl:2002he}
\begin{eqnarray*}
&&q(x,b_{\perp }) =(4\pi)^{N-1}\sum\limits_{n,c}\sum\limits_{a=1}^{N}\int
\left[ \prod\limits_{j=1}^{n}dx_{j} d^{2}r_{\perp j}\right] 
\\
&&\times\delta \left(
1-\sum\limits_{j=1}^{n}x_{j}\right) 
 \delta ^{2}\left(
\sum\limits_{j=1}^{n}x_{j}r_{\perp j}\right)  
 \delta \left( x-x_{a}\right) 
\\
&&\times \delta ^{2}\left( b_{\perp
}+(1-x)r_{\perp a}-\sum\limits_{j\neq a}^{n}x_{j}r_{\perp j}\right)  
\\
&&\times \Psi _{n,c}^{*}(x_{1},\ldots ;r_{\perp 1},\ldots
)\Psi _{n,c}(x_{1},\ldots ;r_{\perp 1},\ldots ),
\end{eqnarray*}
where $a$ denotes the index of the active parton, $N$ is the number of partons in the Fock state
and the sum over $c$ represents the sum over all additional quantum numbers characterizing the Fock state. Here, one explicitly observes $  \lim_{x \to 1}  q(x,b_\perp)  \propto \delta(b_{\perp})$.
Since $H(x,0,t)$ is the Fourier transform of the transverse distribution,
the slope in $-t=\Delta^2_{\perp} $ at the origin measures the rms transverse
radius. As a result, we expect the substantial change in transverse size with $x
$ to be reflected in an equally significant change in slope with $x$. In
particular, as $x \to 1$, the slope should approach zero. Hence, when we
calculate moments of $H(x,0,t)$, the higher the power of $x$, the more
strongly large $x$ is weighted and the smaller the slope should become.
Therefore, this argument makes the qualitative prediction that the slope of
the generalized form factors $A_{n, 0}(t)$ and $\tilde A_{n, 0}(t)$
should decrease with increasing $n$, and we expect that this effect should
be strong enough to be clearly visible in lattice calculations of these form
factors.


Figures~\ref{fig:slopes1} and \ref{fig:slopes2}  show the generalized form factors $A_{n,0}(t)$ and $\tilde A_{n,0}(t)$ for the lowest three moments, $n$ = 1, 2, and 3.
The form factors have been normalized to unity  at $t=0$  to make the dependence of the shape on $n$ more obvious. Note that $A_{1,0}, A_{3,0}$, and $ \tilde A_{2,0}$ depend on the difference between the quark and antiquark distributions 
whereas $ \tilde A_{1,0}, \tilde A_{3,0}$, and $  A_{2,0}$ depend on the sum.
Hence only comparisons between moments differing by $n$ = 2 compare the same physical quantity with different weighting in $x$.
To facilitate determination of the slope of the form factors and to guide the eye, the data have been  fit using a dipole form factor
\begin{equation}
\label{dipole_fit}
A_{n,0}^{\mbox{\footnotesize dipole}}=\frac{a}{(1-\frac{t}{m_d^2})^2}.
\end{equation}
The solid line denotes the least-squares fit and the shaded error band shows
the error in the slope $\Delta m_d$
given by the fit.  Although the dipole fit is purely phenomenological, we note that it is statistically consistent with the lattice data. 

The top panel in Fig.~\ref{fig:slopes1} shows the flavor non-singlet
case $A^u - A^d$, for which the connected diagrams we have calculated
yield the complete answer. It is calculated at the heaviest quark mass we have
considered, corresponding to $m_\pi$ = 0.87 GeV. Note that the form
factors are statistically very well separated, and differ dramatically
for the three moments.   Indeed, as discussed more quantitatively
below, the slope at the origin  decreases  by more than a factor of 2
between $n=1$ and $n=3$, indicating that the transverse size
decreases by more than a factor of 2.  The
top panel of Fig.~\ref{fig:slopes2} shows analogous results for
lighter quarks, $m_\pi$ = 0.74 GeV, where we observe the same
qualitative behavior but slightly weaker dependence on the moment. The
second panel of Fig.~\ref{fig:slopes2} shows the flavor singlet
combination  $A^u + A^d$, for which we have had to omit the
disconnected diagram because of its significantly greater
computational cost. Comparing this figure with the top panel
calculated at the same quark mass, we note that while the connected
contributions to $A^u \pm A^d$ are qualitatively similar, there is
significant quark flavor dependence that can be used to explore the
nucleon wave function. The bottom panel of Fig~\ref{fig:slopes2} shows
the spin-dependent flavor non-singlet form factors $\tilde A^u -
\tilde A^d$ at the heaviest quark mass. Comparison with the top of
Fig.~\ref{fig:slopes1}  displays the difference between  the spin
averaged and spin dependent densities. We observe a striking
difference, in that the change between the $n=1$ and $n=3$ form
factors for  $q(x,b_\perp)_\uparrow - q(x,b_\perp)_\downarrow$ is
roughly 6 times smaller than for $ \frac{1}{2} (q(x,b_\perp)_\uparrow
+ q(x,b_\perp)_\downarrow)$.

Finally, it is useful to use the slope of the form factors at $t$ = 0
to determine the transverse rms radius,
\begin{equation}
\langle r_\perp^2 \rangle^{(n)} = {\frac{\int d^2b_\perp b^2_\perp
    \int dx\, x^{n-1} q(x,b_\perp)}{\int d^2b_\perp  \int dx\, x^{n-1}
    q(x,b_\perp)} }\, .
\end{equation}

Transverse rms radii calculated in this way  for the first three moments are plotted in Fig.~\ref{fig:rms} for $m_\pi$ = 0.870 GeV.   To set the scale, the transverse charge radius at this mass is $ \langle r_\perp^2\rangle_{\mbox{\footnotesize charge}} $ = 0.48 fm, which is two-thirds the experimental transverse size 0.72 fm, reflecting the effect of the absence of a significant pion cloud.  The nonsinglet transverse size $ \langle r_\perp^2\rangle_{\mbox{\footnotesize u-d}} $ = 0.38 fm is slightly smaller than the rms charge radius, but drops 62\% to 0.14 fm for $n$=3. The singlet size $ \langle r_\perp^2\rangle_{\mbox{\footnotesize u+d}} $ is 0.46 fm, and drops 43\% to 0.27 for $n$=3.  This is a truly dramatic change in rms radius arising from changing the weighting by $x^2$. 

To display the change in transverse size with  $x$ in Fig.~\ref{fig:rms}, the
rms radii for each moment are plotted at the average value of $x$ corresponding to that moment. If one neglects the fact that the antiquark distribution contributes to even and odd moments with opposite sign,  the mean value of $x$ in the distributions $q(x) $, $xq(x) $, and $x^2q(x) $ are determined directly from the moments of structure functions measured on the lattice. Applying a small correction for the effect of the alternating antiquark contributions yields the mean values of  $x$ for each moment plotted in the Figure.  The $x$ dependence shown in this Figure is quite striking, with  the nonsinglet transverse size dropping 62\% as 
the mean value of $x$
increases from 0.2 to 0.4, and going to zero when $x$ reaches 1.0.


 
  \section{\label{sec:exploratory-calculation-in-the-chiral-regime}EXPLORATORY CALCULATION IN THE CHIRAL REGIME}
  
Full QCD calculations with light quark masses  are notoriously
expensive, so significant compromises are required to begin to
explore the chiral regime. Our initial exploration of this regime
is a hybrid calculation using MILC configurations
\cite{Bernard:2001av} with staggered sea quarks  and domain wall
valence quarks. The MILC configurations on a $20^3 \times 64$
lattice use strange quark masses $am_s = 0.05$ and   light quark
masses $am_{u+d} = 0.01$ and 0.05 with the
Asqtad action corresponding to lattice spacing
0.13 fm. HYP-smearing~\cite{Hasenfratz:2001hp} with $\alpha_1=0.75,
\alpha_2=0.6,$ and $\alpha_3=0.3$ was used to reduce the effect of
dislocations. Chiral valence quarks were calculated using domain wall
fermions with $L_5$= 16 and $M$ = 1.7.  The lattice size of 2.6 fm can
sustain pions as light as 300 MeV, and our initial calculations were
carried out for pion masses of approximately 343 and 635 MeV.

  In the long term, our plan is to attain high statistics on several lattice volumes and two coupling constants for a range of quark masses. Hybrid partially quenched chiral perturbation theory will be used to correct for the inconsistency between chiral valence quarks and staggered sea quarks, and perturbative renormalization is expected to be adequate because of the improved convergence arising from HYP-smearing.

        \begin{figure}[t]
         \vspace*{-0.7cm}
\hspace*{-0.6cm}  \includegraphics[clip=false,scale=0.30]{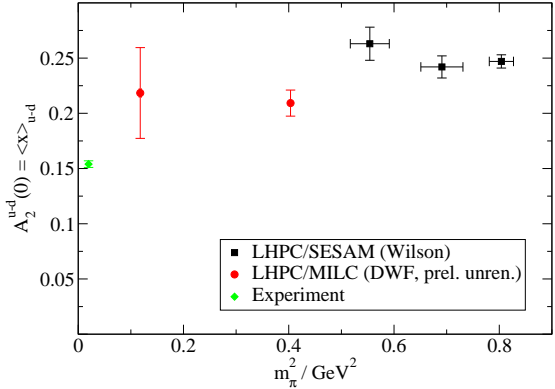}
  \vspace*{-1.0cm}
          \caption{Hybrid chiral calculation of $\langle x \rangle_{u-d}$.}
          \label{fig:MILC1}
          \vspace*{-.5cm}
        \end{figure}

        \begin{figure}[t]
         \vspace*{-0.7cm}
\hspace*{-0.6cm}  \includegraphics[clip=false,scale=0.30]{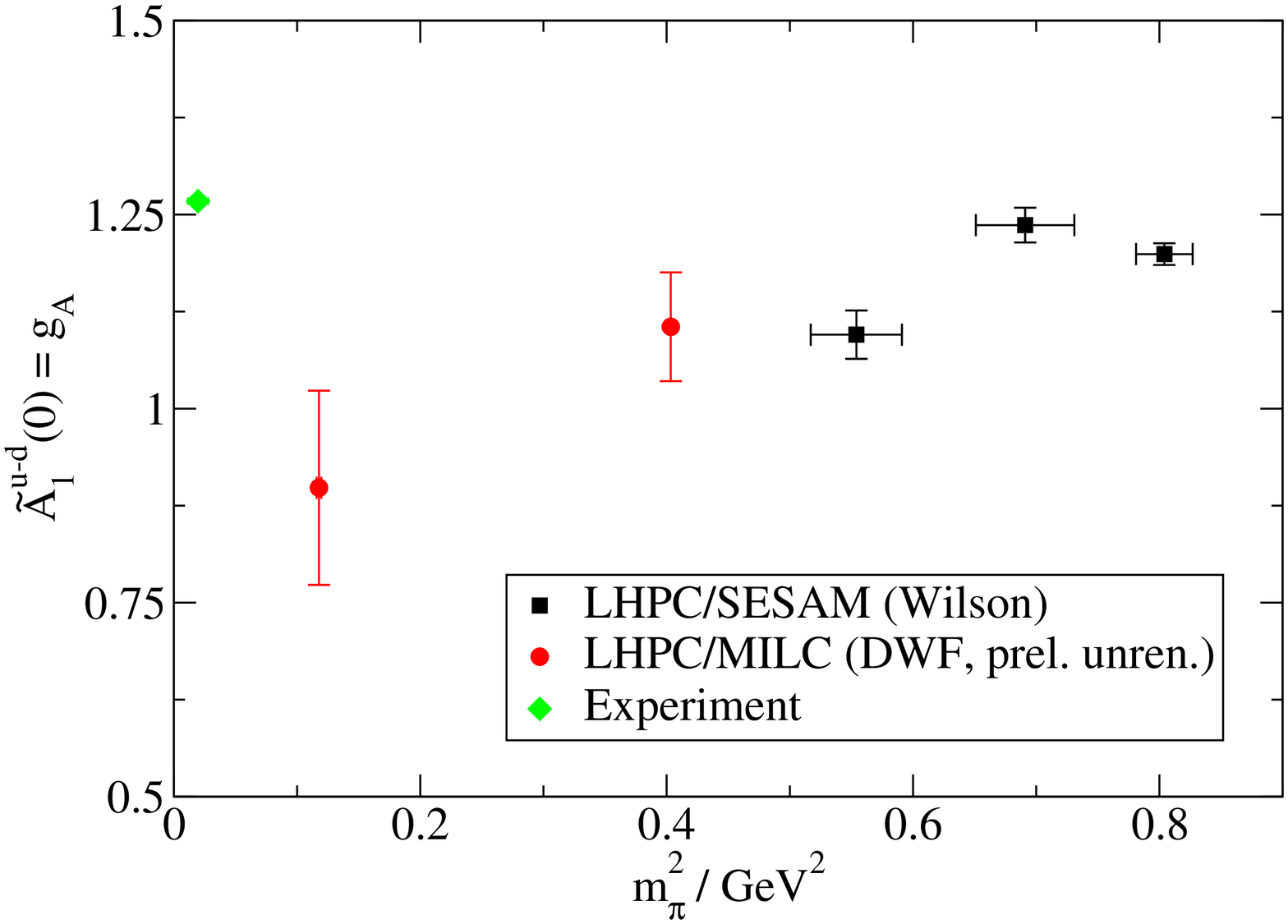}

\vspace*{-0.4cm}
\hspace*{-0.6cm}  \includegraphics[clip=false,scale=0.30]{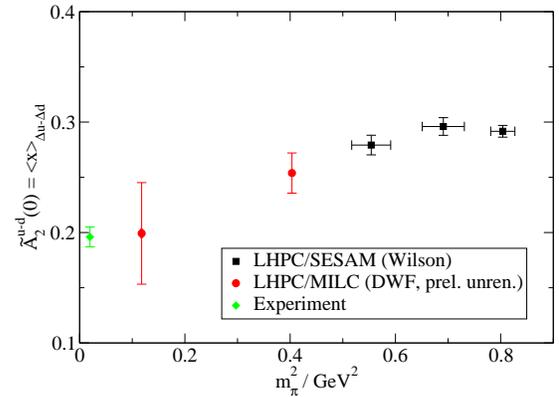}
  \vspace*{-1.0cm}
          \caption{Hybrid chiral calculation of $\langle 1 \rangle_{\Delta u-\Delta d}$
          and $\langle x \rangle_{\Delta u-\Delta d}$.}
          \label{fig:MILC2}
          \vspace*{-.5cm}
        \end{figure}

However, the initial explorations shown in Figs.~\ref{fig:MILC1} and \ref{fig:MILC2} are based on roughly 100 configurations, only have tree-level renormalization, and are not corrected using chiral perturbation theory.
Figure~\ref{fig:MILC1} shows the well-known case of the momentum
fraction, $\langle x \rangle $, for which the chiral extrapolation
formula \cite{Detmold:2001jb} is roughly constant in the heavy pion
regime of the three SESAM points, and then decreases sharply in the
region of $m_{\pi}^2 \sim$ 0.2 GeV$^2$ by 50\% to agree with the
experimental point.  The two MILC points, which unlike the renormalized SESAM results must still be
renormalized, do not yet give any indication of decreasing in the
chiral regime, and it is an open question as to whether the lowest
point is subject to large finite volume corrections.  The axial charge
shown in the upper panel of Fig.~\ref{fig:MILC2} is also presently
unsatisfactory at the light quark point. In this case, it is well
known that finite volume effects produce large discrepancies, so
calculations on a larger lattice are clearly needed in this case. Only
for the case of the first moment of the spin distribution, $\langle x
\rangle_{\Delta u-\Delta d}$, is the qualitative behavior roughly
consistent with experiment.

Whereas these initial results are still far from being quantitatively controlled, they clearly demonstrate feasibility of hybrid calculations in the chiral regime. Improvement of statistics, careful study of finite volume effects, and calculation of hybrid chiral perturbation corrections and renormalization factors offer the potential for a first glimpse at hadron structure in the chiral regime.

  \section{\label{sec:summary-and-outlook}SUMMARY AND OUTLOOK}

  In the heavy pion world presently accessible to unquenched lattice QCD, we have calculated the lowest 3 generalized form factors $A_{n,0}$ and $\tilde A_{n,0}$ to $-t=$ 3 GeV$^2$ and shown that they provide insight into several important aspects of hadron structure. We note that in cases that are comparable, our results are consistent with the calculations of the lowest two moments in Ref.~\cite{Gockeler:2003jf}. The overdetermined method for measuring generalized form factors produces  good statistics up to 3 GeV$^2$, and enables meaningful study of electromagnetic form factors, calculation of the origin of the nucleon spin, and study of the transverse structure of  light cone wave functions. A particularly striking result is the dramatic  62\% decrease in the transverse size $ \langle r_\perp^2\rangle_{\mbox{\footnotesize u-d}} $  between the first and third moment.
We also observed clear dependence of the transverse distribution on flavor and spin and have shown  that the commonly used factorization {\it Ansatz}  $H(x, 0, t) = Q(x)F(t)$ is fundamentally wrong.
    
The most immediate challenges are to extend these calculations to the chiral regime of realistic quark masses,  and to extend techniques for evaluating disconnected diagrams \cite{Neff:2001zr} to these observables. When precise, controlled extrapolations to the physical pion mass are finally achieved,  they will play a special role in our understanding of hadron structure.

  \section{\label{sec:acknowledgments }ACKNOWLEDGMENTS}

We are indebted to members of the MILC collaboration for use of their configurations and for many valuable insights, and to Anna Hasenfratz for helpful discussions.  Computations were performed on the 128-node
Pentium IV cluster at JLab and at ORNL, under the auspices of the
U.S.~DoE's SciDAC initiative. 
P.H.~and W.S.~are grateful for Feodor-Lynen Fellowships from the
Alexander von Humboldt Foundation and thank the Center for
Theoretical Physics at MIT for its hospitality. 
This work is
supported in part by the U.S.~Department of Energy  under
DOE contracts DE-FC02-94ER40818, DE-FG02-91ER40676, and
DE-AC05-84ER40150.  
%
%

\bibliography{GPD_References}

\begin{thebibliography}{10}

\bibitem{Muller:1994fv}
D. M{\"u}ller et~al.,
\newblock
Fortschr. Phys. 42 (1994) 101,
\newblock

\bibitem{Ji:1997ek}
X.D. Ji,
\newblock
Phys. Rev. Lett. 78 (1997) 610,
\newblock

\bibitem{Radyushkin:1997ki}
A.V. Radyushkin,
\newblock
Phys. Rev. D56 (1997) 5524,
\newblock

\bibitem{Diehl:2003ny}
M. Diehl,
\newblock
Phys. Rept. 388 (2003) 41,
\newblock

\bibitem{Hagler:2003jd}
LHPC, P. H{\"a}gler et~al.,
\newblock
Phys. Rev. D68 (2003) 034505,
\newblock

\bibitem{Negele:2003ma}
LHPC, J.W. Negele et~al.,
\newblock (2003),
hep-lat/0309060,
\newblock

\bibitem{Schroers:2003mf}
LHPC, W. Schroers et~al.,
\newblock (2003),
hep-lat/0309065,
\newblock

\bibitem{Haegler:2003pr}
P. H{\"a}gler et~al.,
\newblock In preparation, 2003.

\bibitem{Eicker:1998sy}
TXL, N. Eicker et~al.,
\newblock
Phys. Rev. D59 (1999) 014509,
\newblock

\bibitem{Belitsky:2002kj}
A.V. Belitsky, X.d. Ji and F. Yuan,
\newblock
Phys. Rev. Lett. 91 (2003) 092003,
\newblock

\bibitem{Gayou:2001qd}
Jefferson Lab Hall A, O. Gayou et~al.,
\newblock
Phys. Rev. Lett. 88 (2002) 092301,
\newblock

\bibitem{Dolgov:2002zm}
LHPC, D. Dolgov et~al.,
\newblock
Phys. Rev. D66 (2002) 034506,
\newblock

\bibitem{Gockeler:2003jf}
QCDSF, M. G{\"o}ckeler et~al.,
\newblock (2003),
hep-ph/0304249,
\newblock

\bibitem{Mathur:1999uf}
N. Mathur et~al.,
\newblock
Phys. Rev. D62 (2000) 114504,
\newblock

\bibitem{Burkardt:2000za}
M. Burkardt,
\newblock
Phys. Rev. D62 (2000) 071503,
\newblock

\bibitem{Brodsky:2000xy}
S.J. Brodsky, M. Diehl and D.S. Hwang,
\newblock
Nucl. Phys. B596 (2001) 99,
\newblock

\bibitem{Diehl:2000xz}
M. Diehl et~al.,
\newblock
Nucl. Phys. B596 (2001) 33,
\newblock

\bibitem{Diehl:2002he}
M. Diehl,
\newblock
Eur. Phys. J. C25 (2002) 223,
\newblock

\bibitem{Bernard:2001av}
C.W. Bernard et~al.,
\newblock
Phys. Rev. D64 (2001) 054506,
\newblock

\bibitem{Hasenfratz:2001hp}
A. Hasenfratz and F. Knechtli,
\newblock
Phys. Rev. D64 (2001) 034504,
\newblock

\bibitem{Detmold:2001jb}
W. Detmold et~al.,
\newblock
Phys. Rev. Lett. 87 (2001) 172001,
\newblock

\bibitem{Neff:2001zr}
H. Neff et~al.,
\newblock
Phys. Rev. D64 (2001) 114509,
\newblock

\end{thebibliography}

\end{document}